\documentclass[10pt,letterpaper]{article}
\usepackage[top=0.85in,left=2.75in,footskip=0.75in,marginparwidth=2in]{geometry}

\usepackage[utf8]{inputenc}
\usepackage{multirow}
\usepackage{multicol}
\usepackage{cite}
\usepackage{amsmath}

\usepackage{nameref}
\usepackage{hyperref}
\usepackage{indentfirst}

\usepackage{microtype}
\DisableLigatures[f]{encoding = *, family = * }

\raggedright
\usepackage{changepage}

\usepackage[aboveskip=1pt,labelfont=bf,labelsep=period,singlelinecheck=off]{caption}

\makeatletter
\renewcommand{\@biblabel}[1]{\quad#1.}
\makeatother

\usepackage{lastpage,fancyhdr,graphicx}
\usepackage{epstopdf}
\fancyheadoffset[L]{2.25in}
\fancyfootoffset[L]{2.25in}

\usepackage{color}

\definecolor{Gray}{gray}{.25}

\usepackage{graphicx}

\usepackage{sidecap}

\usepackage{wrapfig}
\usepackage[pscoord]{eso-pic}
\usepackage[fulladjust]{marginnote}
\reversemarginpar

\begin{document}
\vspace*{0.35in}

\begin{flushleft}
{\Large
\textbf\newline{Quantum computational study of chloride ion attack on chloromethane for chemical accuracy and quantum noise effects with UCCSD and k-UpCCGSD ansatzes}
}
\newline
\\
Hocheol Lim\textsuperscript{1,2,3},
Hyeon-Nae Jeon\textsuperscript{3},
June-Koo Rhee\textsuperscript{4},
Byungdu Oh\textsuperscript{5,6,*}, and
Kyoung Tai No\textsuperscript{1,2,5,*}
\\
\bigskip
\bf{\textsuperscript{1}} The Interdisciplinary Graduate Program in Integrative Biotechnology and Translational Medicine, Yonsei University, Incheon, Republic of Korea
\\
\bf{\textsuperscript{2}} Bioinformatics and Molecular Design Research Center (BMDRC), Incheon, Republic of Korea
\\
\bf{\textsuperscript{3}} Department of Biotechnology, Yonsei University, Seoul, Republic of Korea
\\
\bf{\textsuperscript{4}} QuNova Computing, Inc., Daejeon, Republic of Korea
\\
\bf{\textsuperscript{5}} Baobab AiBIO Co., Ltd., Incheon, Republic of Korea
\\
\bf{\textsuperscript{6}} SKKU Advanced Institute of Nanotechnology, Sungkyunkwan University, Suwon, Republic of Korea
\\
\bigskip
* Co-corresponding authors: Byungdu Oh (bdoh@skku.edu) and Kyoung Tai No (ktno@yonsei.ac.kr)

\end{flushleft}

\section*{Abstract}
Quantum computing is expected to play an important role in solving the problem of huge computational costs in various applications by utilizing the collective properties of quantum states, including superposition, interference, and entanglement, to perform computations. Quantum mechanical (QM) methods are candidates for various applications and can provide accurate absolute energy calculations in structure-based methods. QM methods are powerful tools for describing reaction pathways and their potential energy surfaces (PESs). In this study, we applied quantum computing to describe the PES of the bimolecular nucleophilic substitution (S\textsubscript{N}2) reaction between chloromethane and chloride ions. We performed noiseless and noise simulations using quantum algorithms and compared the accuracy and noise effects of the ansatzes. In noiseless simulations, the results from UCCSD and k-UpCCGSD are similar to those of full configurational interaction (FCI) with the same active space, which indicates that quantum algorithms can describe the PES of the S\textsubscript{N}2 reaction. In noise simulations, UCCSD is more susceptible to quantum noise than k-UpCCGSD. Therefore, k-UpCCGSD can serve as an alternative to UCCSD to reduce quantum noisy effects in the noisy intermediate-scale quantum era, and k-UpCCGSD is sufficient to describe the PES of the S\textsubscript{N}2 reaction in this work. The results showed the applicability of quantum computing to the S\textsubscript{N}2 reaction pathway and provided valuable information for structure-based molecular simulations with quantum computing.

\section*{Introduction}
Quantum mechanical (QM) molecular orbital calculation methods can provide accurate absolute energy calculations in structure-based molecular modeling methods, which lead to accurate computational prediction of the binding affinity and selectivity of many drugs to their targets. Although QM methods can provide accurate results quantitatively for intra- and intermolecular interactions in absolute energy calculations, their high computational cost hinders their application to large systems like biomolecules\cite{cite_1}. Owing to practical limitations on computational power, a number of computational methods, including molecular mechanics (MM) force field, hybrid QM/MM, semi-empirical QM, and fragment-based QM methods such as fragment molecular orbital (FMO), have been developed and applied to large biological systems\cite{cite_2, cite_3, cite_4}. One solution to resolve the computational cost of QM methods is quantum computing, and its application to QM methods would help solve numerous biological puzzles with quantum effects.
\\The development of quantum computers has been considered to accelerate dramatically by solving challenging problems that require high computational costs in a number of fields, including physics, chemistry, materials science, and biology\cite{cite_5,cite_6}. Since Feynman proposed the simulation of quantum systems using quantum computers, there have been a number of developments in quantum computing applications\cite{cite_6, cite_7}. Most of the initially proposed quantum algorithms, such as Shor’s, require millions of physical qubits for quantum error correction\cite{cite_8}. However, the currently realized quantum computers are called noisy intermediate-scale quantum (NISQ) devices, whose qubits are in the order of 100, and whose quantum operations are substantially imperfect\cite{cite_9}. Thus, the algorithms for NISQ devices require low circuit depths that allow quantum operation within the limited coherence time of the devices.
\\Quantum chemistry is a promising candidate for applications in quantum computers. It is believed that they will facilitate a more accurate understanding of biological systems, as well as the capability to more accurately simulate a greater range of complex and realistic systems. Quantum computers can provide the full configuration interaction (FCI) for the numerically best wave functions of systems within basis sets using quantum phase estimation algorithms (QPEAs)\cite{cite_10}. However, QPEAs require a number of gates and long circuit depths, which surpass the capabilities of existing NISQ devices.
\\Because of the limitations of the current NISQ system, quantum chemistry algorithms should adopt alternative methods to reduce the quantum errors from NISQ devices. First, most current NISQ algorithms rely on harnessing the power of quantum computers in a hybrid quantum-classical arrangement (HQA). It assigns the classically difficult part of some computations to a quantum computer and performs the classically tractable part on a sufficiently powerful classical device. The variational quantum eigensolver (VQE) is a type of HQA and variational quantum algorithm that variationally updates the parameters of a parameterized quantum circuit\cite{cite_11, cite_12, cite_13}. Second, quantum algorithms are restricted to a subset of degrees of freedom containing essential quantum behavior, called the active space. This approximation allows one to treat more difficult problems by using fewer qubits and lower gate depths. Hartree–Fock (HF)-and/or density functional theory (DFT)-based quantum embedding schemes incorporate a mean field potential generated by the inactive electrons of the environment\cite{cite_14, cite_15}. By adopting these approximations, the applications to quantum chemistry have been expanded from small molecules, such as hydrogen and lithium hydride, to larger molecules, such as oxiranes, and even protein-ligand systems\cite{cite_14, cite_16}.
\\A transition state is the balance point of catalysis in which covalent bonds are partially formed and broken. This is defined as the state with the highest potential energy along the reaction coordinates\cite{cite_17}. The idea of a transition state was developed in 1935 by Eyring, Evans, and Polanyi, who formulated the transition state as a stable state with thermodynamic, kinetic, and statistical-mechanical treatments\cite{cite_18}. The transition states are first-order saddle points on the potential energy surface (PES) in a system of interest, and the geometries can be determined by searching the saddle points and minimum energy pathways on the PES\cite{cite_19, cite_20}. It is feasible to use QM methods to search for geometries on PES.
\\A common example in organic chemistry is the bimolecular nucleophilic substitution reaction (S\textsubscript{N}2). In the S\textsubscript{N}2 reaction mechanism, there is a single transition state in which bond-breaking and bond-making events occur simultaneously. In an example of the S\textsubscript{N}2 reaction between chloromethane (CH\textsubscript{3}Cl) and chloride ion (Cl\textsuperscript{-}), the attack of the nucleophile (Cl\textsuperscript{-}) on the electrophile (CH\textsubscript{3}Cl) forms a transition state in which the carbon under nucleophilic attack has a penta-coordinate. This pushes off the leaving group (Cl\textsuperscript{-}) to the opposite side and forms the product (CH\textsubscript{3}Cl) with inversion of the tetrahedral geometry at the central carbon atom.
\\In this work, we investigated the PES of the S\textsubscript{N}2 reaction between CH\textsubscript{3}Cl and Cl\textsuperscript{-} in the gas phase using VQE and unitary coupled-cluster (UCC) methods. We compared the PES from ansatzes and classical quantum mechanical results in noiseless simulations and compared the quantum noise effects on ansatzes in noise simulations. The intrinsic reaction pathway of the S\textsubscript{N}2 reaction between CH\textsubscript{3}Cl and Cl\textsuperscript{-} was made by 6-31+G*/B3LYP. Based on this path, we performed a single-point analysis with quantum computational algorithms for generating the PES. In quantum algorithms, we used the UCCSD ansatz and the generalized UCC ansatzes (k-UpCCGSD) from one to five products with the HOMO/LUMO active space and Bravyi–Kitaev (BK) transformation. We then performed quantum noise simulations with the UCCSD and k-UpCCGSD ansatzes with a Qulacs-based arbitrary noise model and quantum hardware-driven noise model (IBM’s ibmq-bogota). Therefore, this work demonstrates the applicability of quantum computing to the PES analysis of the S\textsubscript{N}2 reaction between CH\textsubscript{3}Cl and Cl\textsuperscript{-}.
\section*{Methods}
\subsection*{Electronic structure problem of Hamiltonian}
The electronic structure problem involves determining the low-lying energy levels of chemical systems. Based on the Born-Oppenheimer approximation, the nuclei are much heavier than the electrons, meaning they do not move on the same time scale and their behaviors can be decoupled. The energy levels of the electrons in the system are used to solve the non-relativistic time-independent Schrödinger’s equation shown in Eq (\ref{eqn_1}).
\begin{equation}
\label{eqn_1}
    H\textsubscript{el}\vert\Psi\rangle=E\textsubscript{n}\vert\Psi\rangle
\end{equation}

\begin{equation}
\label{eqn_2}
H\textsubscript{el}=-\sum_{i}{\frac{{\bigtriangledown_{r\textsubscript{i}}^{2}}}{ m\textsubscript{e}}}-\sum_{I}\sum_{i}{\frac{Z\textsubscript{I}e\textsuperscript{2}}{\vert R\textsubscript{I}-r\textsubscript{i} \vert}}+\sum_{i}\sum_{j>i}{\frac{e\textsuperscript{2}}{\vert r\textsubscript{i} - r\textsubscript{j} \vert}}
\end{equation}
\noindent The HF method, also called the self-consistent field (SCF) method, approximates that the N-body wave function of the system can be approximated by a single Slater determinant of N spin orbitals, in which each electron evolves in the mean field of the other electrons. The electron Hamiltonian in Eq (\ref{eqn_2}) can be re-expressed based on the solutions of the molecular orbitals as Eq (\ref{eqn_3}).
\begin{equation}\label{eqn_3}
\begin{aligned}
    \widehat{H}\textsubscript{elec}=\sum_{pq}{h\textsubscript{pq}\widehat{a}_{p}^{\dagger}\widehat{a}_{q}+{\frac{1}{2}}}\sum_{pqrs}{h\textsubscript{pqrs}\widehat{a}_{p}^{\dagger}\widehat{a}_{q}^{\dagger}\widehat{a}_{r}\widehat{a}_{s}},\\
    h_{pq}=\int{\phi_{p}^{*}(r)(-{\frac{1}{2}}\bigtriangledown^{2}-\sum_{I}{\frac{Z_{I}}{R_{I}-r}})\phi_{q}(r)dr}, \\
    h_{pqrs}= \int{\frac{\phi_{p}^{*}(r_{1})\phi_{q}^{*}(r_{2})\phi_{r}(r_{2})\phi_{s}(r_{1})}{\vert r_{1}-r_{2} \vert}}
\end{aligned}
\end{equation}

\noindent The molecular orbitals ($\phi$\textsubscript{u}) can contain two electrons, which are spin orbitals of a spin-up ($\alpha$) and spin-down ($\beta$) electron, and can be occupied and virtual. To transform the electronic structure problem into quantum states in quantum computers, the fermionic operators in the Hamiltonian must be transformed into spin operators using mapping methods.

\subsection*{Bravyi-Kitaev Transformation}
Mapping methods from fermionic operators to spin operators have different properties. The Jordan–Wigner (JW) mapper is straightforward, and when using it, qubits can naturally store the occupation of a given spin orbital\cite{cite_21, cite_22}. The parity mapper translates the problem into a representation using parity variables that encode spin variables\cite{cite_23}. While the occupation of a spin orbital is stored locally in the JW, the occupation is stored non-locally in parity. BK encoding is a midway point between the JW and parity encoding methods, in that it compromises the locality of occupation number and parity information\cite{cite_24}. A thorough comparison of BK and JW mappings was performed by Tranter et al\cite{cite_25}. The orbitals store the partial sums of the occupation numbers. The qubit operators for BK encoding are considerably more complicated than those for JW or parity encoding. However, using the BK transform makes the calculations considerably more efficient in many cases. Applying the BK mapping to a fermionic operator results in qubit operations of order \textit{O}(log\textsubscript{2}n) compared to qubit operations of order \textit{O}(n) in the JW mapping.

\subsection*{Variational Quantum Eigensolver (VQE)}
In quantum chemistry, the minimum eigenvalue of a Hermitian matrix for a system is its ground-state energy. The SCF method is an iterative method that involves selecting an approximate Hamiltonian and solving the Schrödinger equation to obtain a more accurate set of orbitals until the results converge. To find the best wave function, the variational method is used to choose the atomic orbital coefficients to minimize the energy, based on the variational principle that the approximate HF wave function is always greater in energy than the exact ground state energy of the system\cite{cite_26}.
\\Because of quantum errors from NISQ devices, most current NISQ algorithms rely on harnessing the power of quantum computers in a hybrid quantum-classical arrangement, where the classically difficult parts are assigned to the quantum computer and the classically tractable parts are calculated on sufficiently powerful classical devices. The variational method in quantum algorithms, referred to as variational quantum algorithms (VQAs), variationally update the parameters of quantum circuits\cite{cite_11,cite_12}. The first proposal of a VQA was the VQE\cite{cite_27,cite_28,cite_29}, originally proposed to solve quantum chemistry problems, and the quantum approximate optimization algorithm\cite{cite_30} was proposed to solve combinatorial optimization problems.

\subsection*{Unitary Coupled-Cluster (UCC) Ansatzes}
The unitary coupled cluster (UCC) method is an extension of the coupled cluster (CC) method, which is one of the most popular post-HF methods\cite{cite_12}. The UCC method creates a parametrized trial state by considering excitations above the initial reference state and can be written as per Eq (\ref{eqn_4}), where the \textit{occ} are occupied orbitals in the reference state, and the \textit{virt} are orbitals that are initially unoccupied in the reference state.
\begin{equation}\label{eqn_4}
    \vert \Psi(\theta)\rangle=U(\theta)\vert\Psi_{HF}\rangle
\end{equation}
\begin{equation}\label{eqn_5}
    U(\theta)=e^{T(\theta)-T(\theta)^{\dagger}}
\end{equation}
\begin{equation}\label{eqn_6}
    T(\theta)=T_{1}(\theta)+T_{2}(\theta)+\cdots
\end{equation}
\begin{equation}\label{eqn_7}
    T_{1}(\theta)=\sum_{{i}\in{occ}, {j}\in{virt}}{\theta_{ji}a_{j}^{\dagger}a_{i}}
\end{equation}
\begin{equation}\label{eqn_8}
    T_{1}(\theta)=\sum_{{i_{1},i_{2}}\in{occ}, {j_{1},j_{2}}\in{virt}}{\theta_{j_{1}j_{2}i_{1}i_{2}}a_{j_{2}}^{\dagger}a_{i_{2}}a_{j_{1}}^{\dagger}a_{i_{1}}}
\end{equation}

\noindent The UCC ansatz can be constructed from the parameterized cluster operator $T(\theta)$ and adds a quantum correlation to the HF ground state $\vert{\Psi_{HF}\rangle}$ \cite{cite_31}. The UCC method is intractable on classical computers but can be efficiently implemented on a quantum computer. It was originally proposed for quantum computational chemistry by Peruzzo et al. and Yung et al\cite{cite_28,cite_32}. The UCC method retains all the advantages of the CC method, with the added benefits of being fully variational and able to converge when using multireference initial states. The UCC ansatz is typically truncated at a given excitation level, usually with single and double excitations, known as UCCSD\cite{cite_12}.
\\In its original form, the UCC ansatz has several drawbacks in its application to larger chemistry problems as well as in other applications. For strongly correlated systems, the widely proposed UCCSD ansatz is expected to have an insufficient overlap with the true ground state and typically results in large circuit depths\cite{cite_33,cite_34}. Heuristic analyses are proposed to mitigate these challenges. For example, the generalized k-UpCCGSD ansatz adopts repeated layers of selected UCC operators\cite{cite_34}. The k-UpCCGSD approach restricts the double excitations to pairwise excitations but allows for k layers of the approach, which offers a good trade-off between accuracy and cost in implantation on quantum computers\cite{cite_34}.

\subsection*{Structure preparation}
Chloromethane (CH\textsubscript{3}Cl) and chloride ion (Cl\textsuperscript{-}) systems for the S\textsubscript{N}2 reaction were prepared using Gaussian 09\cite{cite_35}. We employed density functional theory (DFT) methods with the B3LYP functional and the 6-31+G* basis set in the gas phase. We obtained 23 stationary points using the Hessian matrix calculation of geometries from full optimization for the minimums and for the transition state. In the transition state, we performed vibrational frequency calculations at the same computational level (DFT/B3LYP/6-31+G*) and obtained one imaginary frequency for the transition state (-350.76 cm\textsuperscript{-1}).

\subsection*{Technical details of classical calculations and quantum simulations}
Semi-quantum mechanical calculations (AM1, PM6, and DFTB3) were performed using Gaussian 09\cite{cite_35} and GAMESS\cite{cite_36}. Austin model 1 (AM1) neglects the differential diatomic overlap (NDDO) integral approximation\cite{cite_37}. Parametric method 6 (PM6) is also based on the NDDO approximation and is introduced to correct major errors in AM1 and PM3 calculations\cite{cite_38}. Density functional tight binding (DFTB) is an approximation method of density functional theory (DFT). DFTB3 is a third-order expansion of DFT that improves the Coulomb interaction between atomic partial charges\cite{cite_39}.
\\All traditional QM calculations were performed using PySCF and Psi4\cite{cite_40,cite_41}. In the S\textsubscript{N}2 reaction system (CH\textsubscript{3}Cl and Cl\textsuperscript{-}), we used four calculation levels without active space and two calculation levels with active space as references. Four calculations without active space were performed with HF and full configurational interaction with STO-3G (HF/STO-3G and FCI/STO-3G) and coupled cluster single and double excitation with cc-pVDZ and cc-pV5Z (CCSD/cc-pVDZ and CCSD/cc-pV5Z). The two calculations with active space were performed using CCSD/STO-3G and FCI/STO-3G.
\\All the quantum algorithms were performed using tequila and qiskit\cite{cite_42,cite_43}. We limited the active space to two electrons in four spin orbitals (four qubits), using the BK transformation, “cobyla” optimizer, and UCCSD or k-UpCCGSD ansatz (k=1,2,3,4,5). The error in quantum noise simulations is measured using RMSE, which is a root-mean-square error and a measure of the difference between the two results. First, we used an arbitrary noise model from bit flip, dephasing, depolarizing, two-qubit depolarizing, and amplitude damping noises implemented in Qulacs\cite{cite_44} for the UCCSD and k-UpCCGSD analyses. The final energy value was averaged with 262,144 shots from independent 32 calculations with 8,192 shots per iteration. Second, we used the quantum hardware-driven noise model from IBM’s superconducting transmon quantum computer (\textit{ibmq-bogota}) for the k-UpCCGSD analyses. The final energy value was averaged using 300,000 shots from independent 3 calculations with 100,000 shots per iteration.

\section*{Results}
To apply quantum computing to the reaction pathway, we adopted the zero-sum S\textsubscript{N}2 reaction between chloromethane and chloride, as shown in Figure \ref{Figure1}. We performed single-point calculations to generate PES with a quantum computing framework and compared the results to those obtained from classical quantum mechanics in the S\textsubscript{N}2 reaction.
\subsection*{The S\textsubscript{N}2 reaction of Chloromethane and Chloride ion}
\begin{figure}[ht]

\includegraphics[width=\textwidth]{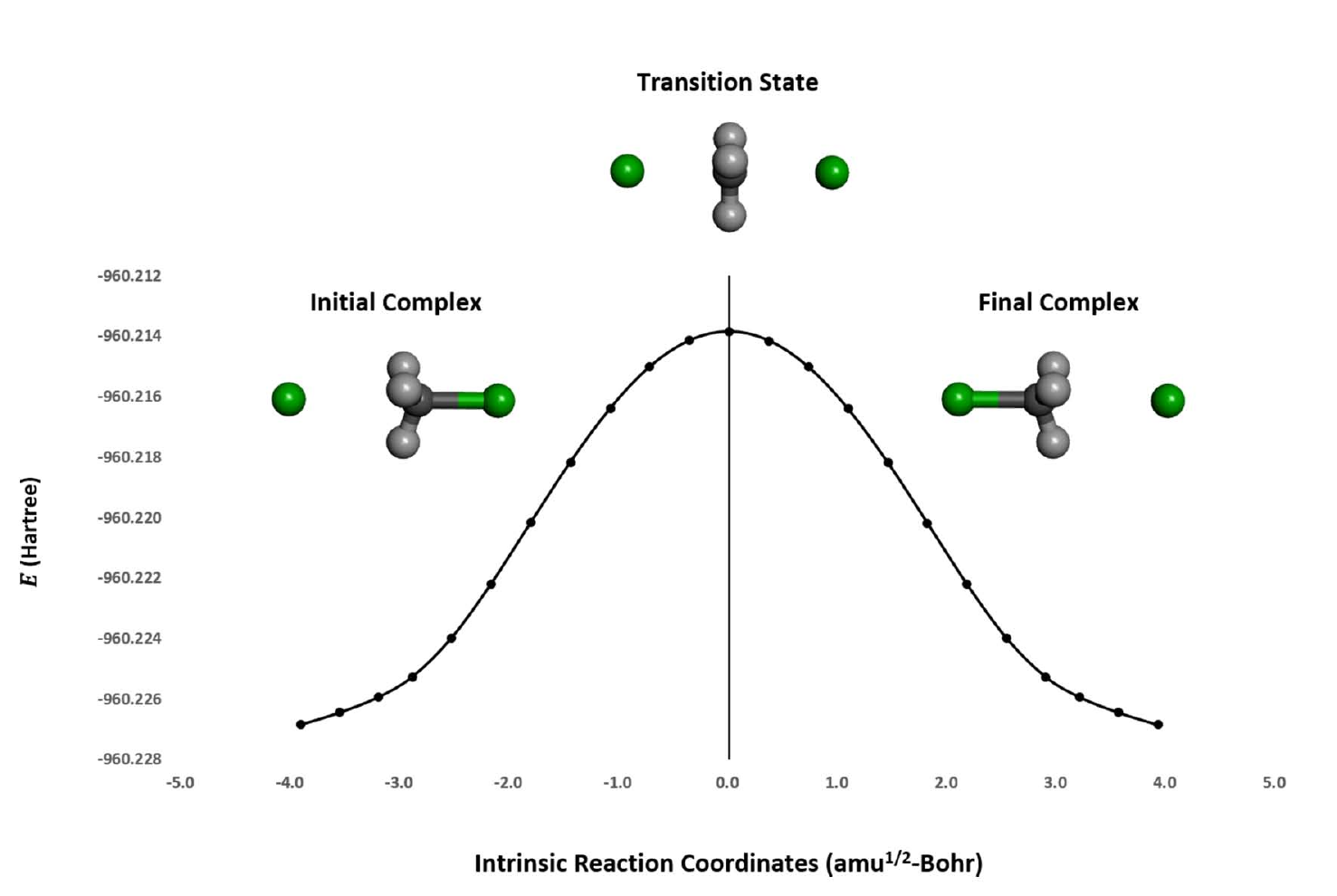}

\caption{\color{Gray} \textbf{The potential energy surface (PES) for the S\textsubscript{N}2 reaction mechanism of CH\textsubscript{3}Cl and Cl\textsuperscript{-} in gas phase.} The carbon atoms are shown in grey, hydrogen atoms are in light grey, and chlorine atoms are in green.}

\label{Figure1} 

\end{figure}

The S\textsubscript{N}2 reaction on the carbon center in the gas phase occurs through a double-well PES from the reactants to the products. In the S\textsubscript{N}2 reaction of chloromethane and chloride, we created an intrinsic reaction pathway of 23 steps in the gas phase, from the initial complex to the final complex, using classical quantum mechanics and B3LYP/6-31+G* (Figure \ref{Figure1}). The energy diagram of the transition state is shown in \nameref{Figure S1}.
\\Semi-QM methods can be one of the alternatives for finding intrinsic reaction pathways. Based on the pathway constructed by the ab initio QM method, we performed single-point calculations in the gas phase for analysis with semi-QM methods (AM1, PM6, and DFTB3), the results of which are shown in \nameref{Figure S2}. The results from the AM1 and PM6 methods appear graphically as a bell-shape curve and were closer to those from B3LYP/6-31+G* than to those from DFTB3 in the S\textsubscript{N}2 reaction system of chloromethane and chloride ions.

\subsection*{Quantum Noiseless Simulations of UCCSD and k-UpCCGSD Ansatzes and Comparison with Classical Results}
\begin{figure}[ht]

\includegraphics[width=\textwidth]{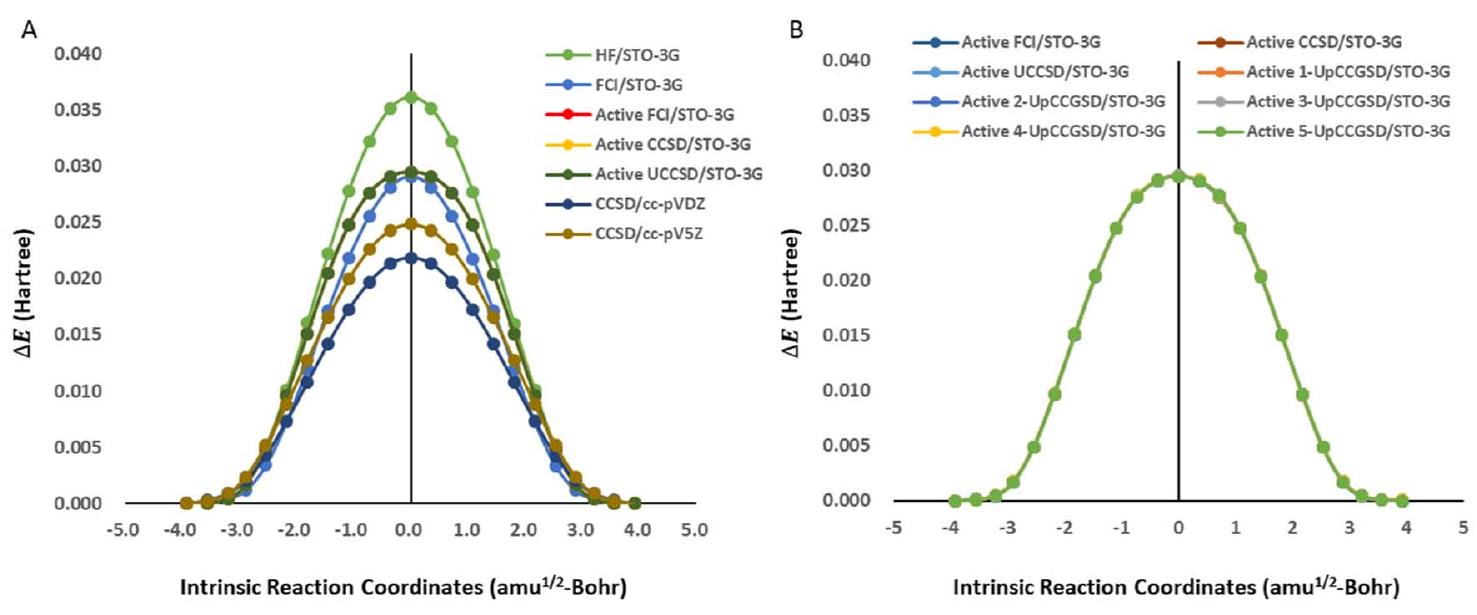}

\caption{\color{Gray} \textbf{Potential energy surfaces from noiseless quantum simulations.} (A) Comparison of UCCSD ansatz and classical quantum calculations, which includes the HF/STO-3G, FCI/STO-3G, CCSD/cc-pVDZ and CCSD/cc-pV5Z without active space, the FCI/STO-3G and CCSD/STO-3G with HOMO/LUMO active space, and UCCSD/STO-3G with active space. (B) Comparison of UCCSD ansatz and generalized k-UpCCGSD ansatzes with the classical results of the CCSD/STO-3G and FCI/STO-3G with active space.}
\label{Figure2} 
\end{figure}

To establish a baseline for classical quantum mechanics, we performed a series of calculations, as shown in Figure \ref{Figure2}A. In classical calculations without an active space, we adopted the HF/STO-3G, FCI/STO-3G, CCSD/cc-pVDZ, and CCSD/cc-pV5Z levels. In classical calculations with the HOMO/LUMO active space, we adopted the CCSD/STO-3G and FCI/STO-3G levels. To obtain the PES in the S\textsubscript{N}2 reaction pathway with quantum computing algorithms, we searched for the ground-state energy of their molecular Hamiltonians using the HOMO/LUMO active space, UCCSD ansatz, STO-3G basis set, and BK transformation (UCCSD/STO-3G). We estimated the relative energy differences between the initial point and all points in the pathway.

To compare the chemical accuracy of the quantum computing results, we measured the RMSE of the PES from the quantum algorithms and classical results, which are shown in Figure \ref{Figure2}A and summarized in Table \ref{Table 1}. First, we compared UCCSD/STO-3G with an active space and other classical results without an active space. The RMSE values of HF/STO-3G or FCI/STO-3G without active space are 1.77 and 1.20 kcal/mol, respectively. The results from the UCCSD/STO-3G with active space are similar to those from FCI/STO-3G without active space but cannot come close to those with chemical accuracy (less than 1 kcal/mol). The RMSE values of CCSD/cc-pVDZ and CCSD/cc-pV5Z without active space are 3.05 and 1.89 kcal/mol, respectively. UCCSD/STO-3G, with active space, cannot overcome the basis set limit and the error from the active space approximation within the chemical accuracy. Second, we compared UCCSD/STO-3G with other classical results with the same active space. The RMSE values of CCSD/STO-3G and FCI/STO-3G with active space were 7.63×10\textsuperscript{-4} and 7.84×10\textsuperscript{-4} kcal/mol, respectively. The results from UCCSD/STO-3G are close to those from FCI/STO-3G or CCSD/STO-3G, with an active space within chemical accuracy (under 1 kcal/mol). Third, we compared k-UpCCGSD with other classical results. In comparison with the classical results without an active space, k-UpCCGSD showed results similar to those of UCCSD. Compared to the classical result with active space, k-UpCCGSD showed more inaccurate results than UCCSD, where the RMSE values between k-UpCCGSD and FCI/STO-3G with active space are 2.10×10\textsuperscript{-2} and 3.20×10\textsuperscript{-2} kcal/mol, respectively. Therefore, k-UpCCGSD can describe the PES of the S\textsubscript{N}2 reaction with chemical accuracy in noiseless simulations.

To compare the effects of different ansatzes in quantum computing algorithms, we measured the RMSE between the UCCSD/STO-3G and k-UpCCGSD ansatz (k=1,2,3,4,5) with active space, as shown in Figure \ref{Figure2}B and summarized in \nameref{Table S1}. All RMSE values between the ansatzes were chemically accurate. The RMSE values between k-UpCCGSD and UCCSD were 2.87×10\textsuperscript{-2}, 2.54×10\textsuperscript{-2}, 3.17×10\textsuperscript{-2}, 1.98×10\textsuperscript{-2}, and 2.07×10\textsuperscript{-2} kcal/mol from 1-UpCCGSD to 5-UpCCGSD, respectively. The RMSE values for UCCSD/STO-3G showed a decreasing tendency as the k value increased. Therefore, the higher k products of k-UpCCGSD provided more accurate results for the S\textsubscript{N}2 reaction.

\begin{table}[h!]
\begin{adjustwidth}{-1.0in}{0in}
\begin{center}
\caption{\bf RMSE of ansatzes with classical results in noiseless simulations (kcal/mol).}
\begin{tabular}{|c|c|c|c|c|c|c|}
\hline
\centering
\textbf{Active Space} & \textbf{Full} & \textbf{Full} & \textbf{Full} & \textbf{Full} & \textbf{Active Space} & \textbf{Active Space}\\ 
\textbf{Ansatz} & \textbf{HF} & \textbf{FCI} & \textbf{CCSD} & \textbf{CCSD} & \textbf{CCSD} & \textbf{FCI} \\
\textbf{STO-3G} & \textbf{STO-3G} & \textbf{STO-3G} & \textbf{cc-pVDZ} & \textbf{cc-pV5Z} & \textbf{STO-3G} & \textbf{STO-3G} \\ 
\hline
    UCCSD & 1.77 & 1.20 & 3.05 & 1.89 & $7.63\times10^{-4}$ & $7.84\times10^{-4}$\\ \hline
    1-UpCCGSD & 1.77 & 1.21 & 3.06 & 1.90 & $2.91\times10^{-2}$ & $2.93\times10^{-2}$\\ \hline
    2-UpCCGSD & 1.77 & 1.20 & 3.05 & 1.89 & $2.54\times10^{-2}$ & $2.55\times10^{-2}$\\ \hline
    3-UpCCGSD & 1.76 & 1.21 & 3.06 & 1.90 & $3.18\times10^{-2}$ & $3.20\times10^{-2}$\\ \hline
    4-UpCCGSD & 1.76 & 1.21 & 3.06 & 1.90 & $2.01\times10^{-2}$ & $2.04\times10^{-2}$\\ \hline
    5-UpCCGSD & 1.77 & 1.21 & 3.05 & 1.89 & $2.09\times10^{-2}$ & $2.10\times10^{-2}$\\ \hline
\end{tabular}
\label{Table 1}
\end{center}
\end{adjustwidth}
\end{table}

\begin{figure}[ht]
\includegraphics[width=\textwidth]{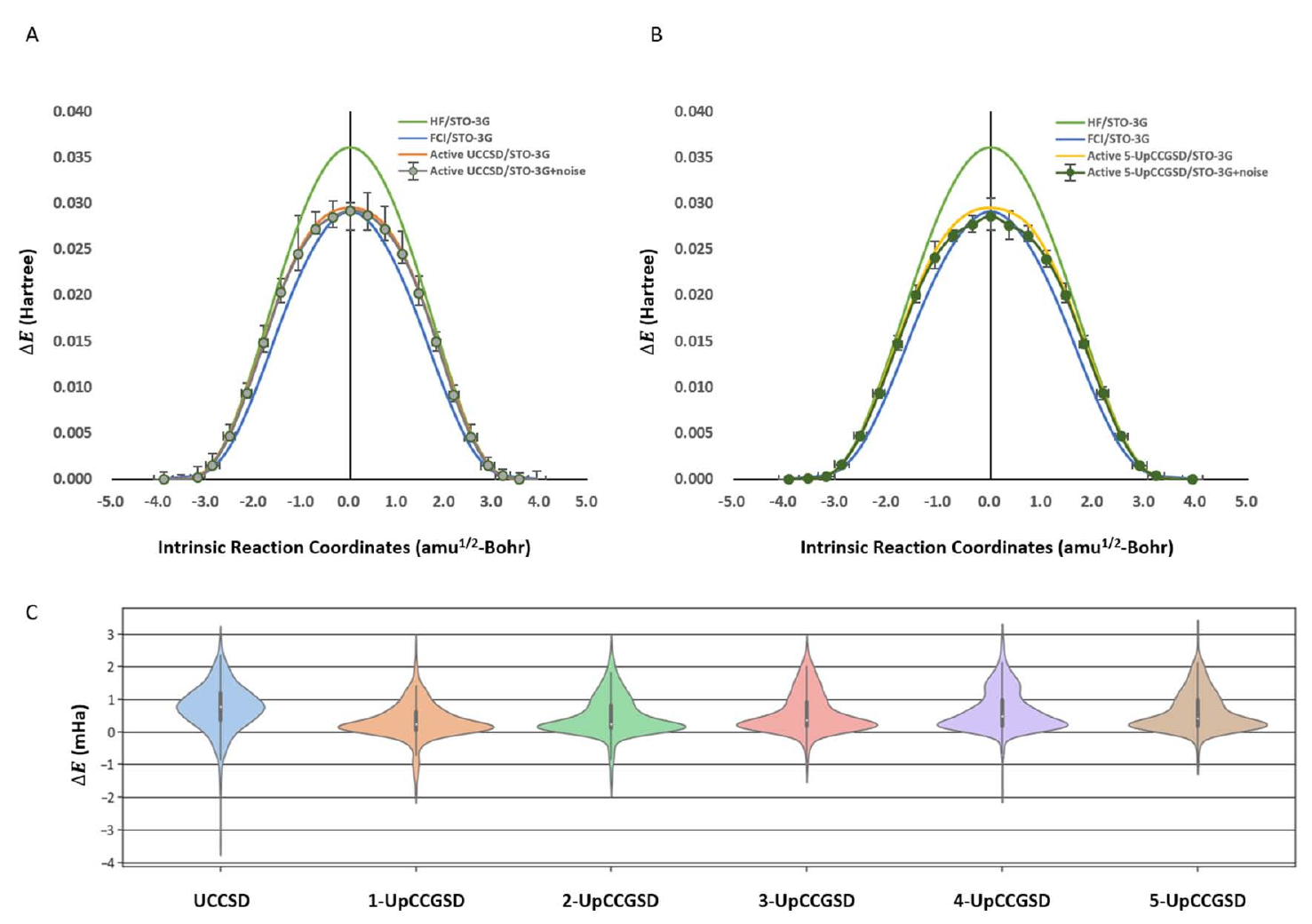}
\caption{\color{Gray} \textbf{Potential energy surfaces from quantum noise simulations with the Qulacs-based arbitrary noise model} (A, B) Quantum noisy effects on UCCSD and 5-UpCCGSD ansatzes with STO-3G. The classical QM results of HF/STO-3G and FCI/STO-3G without active space are shown in green and blue, respectively. The noiseless results from UCCSD ansatz and 5-UpCCGSD ansatz with HOMO/LUMO active space are shown in orange and yellow, respectively. The shot-based quantum noise simulation results from the UCCSD ansatz and 5-UpCCGSD ansatz with Qulacs quantum noises are shown in grey and dark green. The error bar is based on 262,144 shots, from 32 times 8,192 shots, in quantum noises and is shown in black line. (C) Error in energy in ansatzes with HOMO/LUMO active space and STO-3G. The error was calculated from the difference between the noiseless results and noisy results of each ansatz in all 23 steps.}
\label{Figure3}
\end{figure}
\subsection*{Quantum Noise Simulations of UCCSD and k-UpCCGSD Ansatzes with Qulacs-based Arbitrary Noise Model}
Near-term quantum devices are noise-prone, so an explanation of how a VQE is calculated in the presence of quantum noise is important to evaluate the performance of quantum algorithms. To simulate the VQE calculations with quantum noise, we performed quantum noise simulations with the Qulacs-based arbitrary noise model, where we used 262,144 shots from independent 32 calculations with 8,192 shots per iteration. The PES from UCCSD/STO-3G and 5-UpCCGSD/STO-3G is shown in Figure \ref{Figure3}, and those from the other 1,2,3,4-UpCCGSD/STO-3G are shown in \nameref{Figure S3}. In the quantum noise simulations of UCCSD/STO-3G, the upper bound on the error was 1.54 mHa on average, while the lower bound on the error was 1.26 mHa on average of the PES. When the k product increased from 1 to 5 in the quantum noise simulations from the generalized k-UpCCGSD analyses, the upper bounds on the error were 0.889, 0.859, 0.752, 0.696, and 0.721 mHa, respectively, and the lower bounds on the error were 0.690, 0.643, 0.647, 0.708, and 0.589 mHa, respectively. In quantum noise simulations with the Qulacs-based arbitrary noise model, the quantum noise errors from k-UpCCGSD/STO-3G were smaller than those from UCCSD/STO-3G.

\begin{figure}[ht]
\includegraphics[width=\textwidth]{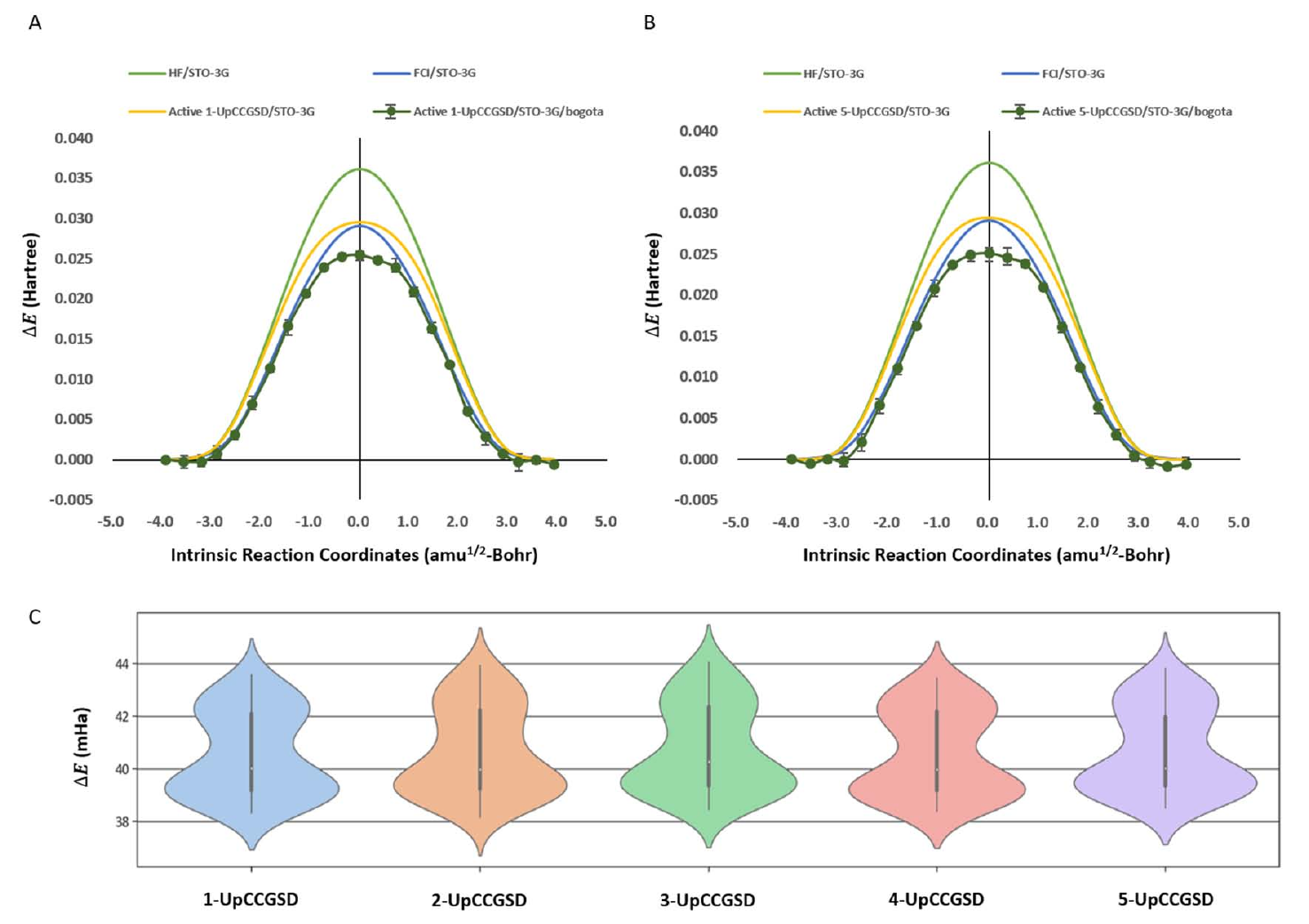}
\caption{\color{Gray} \textbf{Potential energy surfaces from quantum noise simulations with the \textit{ibmq-bogota} noise model.} (A, B) Quantum noisy effects on 1-UpCCGSD and 5-UpCCGSD ansatzes with STO-3G. The classical QM results of HF/STO-3G and FCI/STO-3G without active space are shown in green and blue, respectively. The noiseless results are shown in yellow. The shot-based quantum noise simulation results from 1-UpCCGSD ansatz and 5-UpCCGSD ansatz with \textit{ibmq-bogota} noise models are shown in dark green. The error bar is based on 300,000 shots from 3 times 100,000 shots in quantum noises and is shown in black line. (C) Error in energy in ansatzes with HOMO/LUMO active space and STO-3G. The error was calculated from the difference between the noiseless results and noisy results of each ansatz in all 23 steps.}
\label{Figure4}
\end{figure}

\subsection*{Quantum Noise Simulations of k-UpCCGSD Ansatzes with Hardware-driven Noise Models}
To simulate the VQE calculations with quantum noises from the NISQ devices, we performed quantum noise simulations with 300,000 shots from three 100,000 shots using qasm simulators with the hardware-driven noise model (\textit{ibmq-bogota}). The PES from 1-UpCCGSD and 5-UpCCGSD with the \textit{ibmq-bogota}-driven noise model are shown in Figure \ref{Figure4}A and \ref{Figure4}B, and the errors of the k-UpCCGSD ansatzes are shown in Figure 4C. The PES with the \textit{ibmq-bogota}-driven noise model manifested as a bell-shape curve and was found to follow the PES from the FCI/STO-3G without active space or noiseless simulation results. However, quantum noise resulted in an average error of 40.654 mHa in k-UpCCGSD compared to noiseless results. The standard deviation in the three independent simulations was 0.436 mHa, which was consistent with the results of the three simulations. Therefore, the relative PES from the hardware-driven noise model is useful, but the absolute energy values have quite high errors compared with the noiseless results.


\section*{Discussion}
The generalized k-UpCCGSD was devised for a trade-off between accuracy and cost.\cite{cite_34} In the noiseless quantum simulation of the S\textsubscript{N}2 reaction, UCCSD was found to be more accurate than the k-UpCCGSD, where both UCCSD and k-UpCCGSD are also chemically accurate compared to FCI/STO-3G with the same active space. However, UCCSD showed a wider variance than k-UpCCGSD in Qulacs-based arbitrary noise simulations, which indicates that UCCSD is more susceptible to quantum noise than k-UpCCGSD. In comparison with k-UpCCGSD, the higher k products made the circuit depths longer and the ansatzes more susceptible to quantum noise. Interestingly, there was no clear tendency related to the k products and errors in the Qulacs-based noise simulations, and the ibmq-bogota-driven noise simulation. This may be because the circuit depth of k-UpCCGSD in the S\textsubscript{N}2 reaction is so long that the quantum noise is saturated. 1-UpCCGSD has the shortest circuit depth in k-UpCCGSD, and its qiskit code is shown in the Supporting Information, but it showed a similar accuracy to the ansatzes with higher k products in this work. Although 1-UpCCGSD is sufficient to describe the PES with chemical accuracy in noiseless simulations in this work, 1-UpCCGSD has been reported to work with less accuracy than UCCSD and higher k-UpCCGSD in some cases\cite{cite_34,cite_45}. This is because the HOMO/LUMO active space approximation reduces the single and double excitations of orbitals, and also decreases the accuracy improvement effects of the higher-k products. Therefore, k-UpCCGSD can serve as an alternative analysis of UCCSD to reduce quantum noisy effects in the NISQ era, and k-UpCCGSD is sufficient to describe the PES of the S\textsubscript{N}2 reaction in this work.
\\Interestingly, the error of the hardware-driven noise simulations was high, but the standard deviations of the three independent simulations were small. In the Qulacs-based noise simulations, quantum noises were randomly generated, where the higher samples per iteration and the higher number of independent simulations could deal with the quantum noise and reduce the errors with noiseless results. However, the error from the \textit{ibmq-bogota}-driven noise simulation was not reduced by 100,000 samples per iteration or by three independent simulations. The three independent simulations showed very similar results, with a standard deviation of 0.436 mHa. This may indicate that new error mitigation algorithms are required to reduce the errors from quantum noises of NISQ devices, not only for the higher samples, but also for the more independent simulations.
\\The transition state theory explains the reaction rates of chemical and enzymatic reactions. Computational analysis of the reaction pathway is important in the molecular design of transition-state analogs, understanding chemical and enzymatic reactions, and enzyme modification for higher activity. In this study, we used the ab initio QM (6-31+G*/B3LYP) method to construct an intrinsic reaction pathway. However, searching the intrinsic reaction pathway with the ab initio QM method would require a high computational cost in larger systems such as enzymes. To reduce the computational cost of finding the reaction pathway, it is necessary to use alternative approximate methods, including the semi-empirical QM, QM/MM, and FMO methods, in larger systems\cite{cite_20,cite_46}. The application of quantum computing to search the reaction pathway is promising for the acceleration and accuracy of computational analysis.
\\In this study, we analyzed the potential energy surfaces of the S\textsubscript{N}2 reaction using quantum computational algorithms by making simplifications with active space approximations. Thus, we can select a subset of molecular orbitals for quantum mechanical calculations using near-term quantum computers. The quality of the electronic structure predictions depends on a small subset of frontier orbitals; therefore, the active space scheme can be a solution to interesting quantum chemistry problems with NISQ hardware and restricted qubit numbers. As the NISQ hardware evolves and availability of quantum qubits increases, it would be possible with quantum computing algorithms to calculate single point analysis without active space approximations and with the higher basis sets for more accurate results. Furthermore, quantum computational applications can be expanded to search for intrinsic reaction paths in the reaction mechanisms. Major bottlenecks in the application of quantum computing may be the restricted qubit numbers and the development of ansatzes for quantum chemistry calculations that are intractable to NISQ machines with many qubits. As quantum computers and ansatzes advance, quantum computational applications to quantum chemistry problems are expected to be solutions to important problems in physics, chemistry, biology, and medicine.

\section*{Conclusion}
Quantum computers are considered to dramatically accelerate collection of information on the collective properties of quantum states. Quantum mechanical methods are useful applications in quantum computers. Structure-based molecular simulations using quantum mechanics and quantum computers have accelerated structure-based drug discovery. As a case study for structure-based molecular simulations, we investigated the potential energy surfaces of the S\textsubscript{N}2 reaction pathway using quantum computing. The results shed light on the applicability of quantum computing to structure-based molecular simulations. If the system is sufficiently simplified and errors are handled using appropriate error-mitigation techniques, the NISQ hardware is capable of reproducing the trend obtained from classical computations. We look forward to seeing how quantum computation will accelerate structure-based molecular simulations and yield important impacts on structure-based drug discovery.


\section*{Supporting Information}
Supproting information includes Abbreviations, Figure S1, Figure S2, Table S1, and qiskit code for 1-UpCCGSD in this work.

\subsection*{Figure S1}
\label{Figure S1}
Energy diagram of the system of chloride ion attack on chloromethane.
\subsection*{Figure S2}
\label{Figure S2}
The potential energy surface (PES) from semi-quantum mechanical methods.
\subsection*{Figure S3}
\label{Figure S3}
Quantum noisy effects on k-UpCCGSD ansatzes with STO-3G and HOMO/LUMO active space.
\subsection*{Table S1}
\label{Table S1}
The RMSE between UCCSD and k-UpCCGSD in noiseless simulations (kcal/mol)


\section*{Acknowledgments}
This research was financially supported by the Ministry of Trade, Industry, and Energy (MOTIE), Korea, under the “Infrastructure Support Program for Industry Innovation” (reference number P0014714) supervised by the Korea Institute for Advancement of Technology (KIAT). This research was supported by the quantum computing technology development program of the National Research Foundation of Korea (Grant No. 2020M3H3A111036512 and Grant No. 2019M3E4A1080227).

\section*{Author contributions}
H.L. conceptualized this study. H.J. supported construction of the S\textsubscript{N}2 reaction pathway. B.O. and K.N. advised on the study.

\section*{Competing interests}
The authors declare no competing interests.

\bibliography{library}

\bibliographystyle{unsrt}

\end{document}